\documentclass[prl,twocolumn,aps,floatfix,showpacs,tightenlines,showpacs,superscriptaddress]{revtex4}
\usepackage{amssymb,amsmath,amsthm,amsbsy,epsfig,color,graphicx}

\newcommand{\beq}{\begin{eqnarray}}
\newcommand{\eeq}{\end{eqnarray}}
\newcommand{\be}{\begin{equation}}
\newcommand{\ee}{\end{equation}}
\newcommand{\ket}[1]{ \mid#1\rangle}
\newcommand{\bra}[1]{\langle#1\mid}
\newcommand{\Tr}{\mathrm{Tr}}

\begin{document}

\title{Scaling of the R\'enyi entropies in gapped quantum spin systems:\\
       Entanglement-driven order beyond symmetry breaking}
\author{S. M. Giampaolo}
\affiliation{Dipartimento di Ingegneria Industriale, Universit\`a degli Studi di Salerno,
Via Ponte don Melillo, I-84084 Fisciano (SA), Italy}
\affiliation{CNISM Unit\`a di Salerno, I-84084 Fisciano (SA), Italy}

\author{S. Montangero}
\affiliation{Institut f\"{u}r Quanteninformationsverarbeitung,
Universit\"{a}t Ulm, D-89069 Ulm, Germany}

\author{F. Dell'Anno}
\affiliation{Dipartimento di Ingegneria Industriale, Universit\`a degli Studi di Salerno,
Via Ponte don Melillo, I-84084 Fisciano (SA), Italy}
\affiliation{CNISM Unit\`a di Salerno, I-84084 Fisciano (SA), Italy}


\author{S. De Siena}
\affiliation{Dipartimento di Ingegneria Industriale, Universit\`a degli Studi di Salerno,
Via Ponte don Melillo, I-84084 Fisciano (SA), Italy}
\affiliation{CNISM Unit\`a di Salerno, I-84084 Fisciano (SA), Italy}

\author{F. Illuminati}
\thanks{Corresponding author: illuminati@sa.infn.it}
\affiliation{Dipartimento di Ingegneria Industriale, Universit\`a degli Studi di Salerno,
Via Ponte don Melillo, I-84084 Fisciano (SA), Italy}
\affiliation{CNISM Unit\`a di Salerno, I-84084 Fisciano (SA), Italy}

\date{August 2, 2012}


\begin{abstract}
We investigate the scaling of the R\'enyi $\alpha$-entropies in one-dimensional gapped quantum spin models.
We show that the block entropies with $\alpha > 2$ violate the area law monotonicity
and exhibit damped oscillations. Depending on the existence of a factorized ground state, the oscillatory behavior occurs either below factorization or it extends indefinitely. The anomalous scaling corresponds to an entanglement-driven order that is independent of ground-state degeneracy and is revealed by a nonlocal order parameter defined as the sum of the single-copy entanglement over all blocks.
\end{abstract}


\maketitle

During the past decade the use of entanglement in the study of complex quantum systems has developed at an increasingly fast
pace~\cite{AmicoFazioOsterlohVedral2008,EisertPlenioCramer2010,CalabreseCardyDoyon2009,Ladd2010}.
Extended analyses have established the monotonic scaling of the von Neumann entropy in the ground state of one-dimensional spin models~\cite{VidalLatorreRicoKitaev2003,LatorreRicoVidal2004}, and its profound relations with conformal field theory (CFT) in the gapless cases~\cite{HolzheyLarsenWilczek1994,CalabreseCardy2004,CalabreseCardy2009}.
Much attention is currently being devoted to the R\'enyi entropies (RE), whose infinite hierarchy provides the most complete information on the spectrum of the reduced density matrix. The general characterization of quantum states provided by the RE plays a relevant role in determining the scaling properties of numerical algorithms based on matrix product
states~\cite{Shuchetal2008,Tagliacozzoetal2008,Pollmannetal2009,VerstraeteCirac2009}. The RE are also a useful tool to determine the continuous or discontinuous nature of a phase transition~\cite{ErcolessiEvangelistiFranchiniRavanini2011} and to estimate quasi-long-range order in low-dimensional systems~\cite{DalmonteErcolessiTaddia2011}. Furthermore, the concept of topological entanglement entropy \cite{Kitaev2006,Levin2006} can be extended to the RE for which it has been shown to coincide with the total quantum dimension~\cite{Flammiaetal2009}.

Above all, there is a growing awareness that the entanglement properties of quantum ground states provide the most fundamental characterization of quantum phases of matter beyond the traditional approach, borrowed from classical statistical mechanics, based on symmetry breaking and local order parameters \cite{Balents2012}. In this perspective, the RE and their topological components are being actively investigated in various problems at the cutting edge of condensed matter physics, including Bose-Hubbard spin liquids \cite{Isakov2011}, frustrated models on nontrivial lattice geometries \cite{Vishwanath2011}, non-Abelian fractional Hall systems~\cite{LiHaldane2008}, and low-dimensional gapless models \cite{ThomaleArovasBernevig2010,Hastings2012}.
For a bipartite quantum system in a global pure state $\rho_{AB} = \ket{\Psi_{AB}}\bra{\Psi_{AB}}$, the RE of subsystem $A$ (and analogously for subsystem $B$) are defined as:
\begin{equation}
S_{\alpha} (A) = \frac{1}{1-\alpha} \ln \left[ \Tr \left( \rho_{A}^{\alpha} \right) \right] \; ,
\label{Renyi}
\end{equation}
where $\rho_A = \Tr_B \rho_{AB}$ is the reduced density matrix of subsystem $A$, and the parameter $\alpha$ takes nonnegative real values. The von Neumann entropy $S_1$ is
recovered in the limit $\alpha \rightarrow 1^{+}$.
In the opposite limit, the entanglement entropy $S_{\infty} = - \ln \left(\lambda_{max}\right)$, where $\lambda_{max}$ is the largest eigenvalue of $\rho_A$, and it coincides with the single-copy entanglement~\cite{CramerEisert}.
%
When the RE $S_{\alpha} (A)$ are computed for blocks $A$ of $\ell$ spins in the ground state of gapless models, 
it is observed that, 
for $\alpha > 2$, they are characterized by large sub-leading corrections that violate monotonicity
in $\ell$~\cite{CardyCalabrese2010,CalabreseCampostrinietal2010,HastingsGonzalezKallinMelko2010}. 
Investigations on the nature of these corrections have unveiled the role played by the parity of the number of spins forming the block~\cite{SongRachelLeHur2010,XavierAlcarez2011,FagottiCalabrese2011}. The significance of these findings stimulates the investigation of the scaling behavior of RE in general gapped systems.

In this work we present a comprehensive study of the ground-state RE
for one-dimensional gapped quantum spin models in external field $h$.
We show that for models admitting ground-state factorization there exist two different scaling regimes, separated by the factorizing field $h_f$. In the region $h < h_f$ all the RE of order $\alpha > 2$ exhibit a non monotonic scaling with damped oscillations in $\ell$. These oscillations are due to a series of crossovers between the eigenvectors of the reduced block density matrix associated to its largest eigenvalue, and thus to the bipartite ground-state geometric block entanglement. In all models admitting factorization, the area law behavior of all RE is restored for $h > h_f$. For models that do not admit a factorization point, the oscillatory behavior persists indefinitely, even beyond the critical point. These phenomena are associated to an entanglement-induced order revealed by a global order parameter defined as the sum of the single-copy entanglement over all block sizes. This entanglement-driven ordered pattern is independent of the Ginzburg-Landau magnetic order and superimposed on it in a symmetry broken phase.

To set the stage, let us first consider the class of translationally invariant, one-dimensional $XY$ spin models
\begin{equation}\label{Hamiltonian1}
H_{xy} = \frac{1}{2}\sum_{i} (1+\gamma) \sigma_{i}^x \sigma_{i+1}^x+ (1-\gamma) \sigma_{i}^y \sigma_{i+1}^y - h\sum_{i} \sigma_{i}^z \, ,
\end{equation}
where $\sigma_{i}^\alpha$ $(\alpha=x,y,z)$ stands for the spin-$1/2$ Pauli operator on site $i$, $h$
is the external transverse field, and $\gamma$ is the anisotropy, taking values in the interval $[0,1]$, whose extremes correspond, respectively, to the gapless isotropic $XX$ model and to the Ising case. In the thermodynamic limit, at $h_{c}=1$ the system undergoes a quantum phase transition, developing a nonvanishing spontaneous magnetization $M_x$. In the ordered phase the ground state $\ket{G}$ is twofold degenerate and the energy spectrum is gapped.
For the $XX$ model the spectrum is gapless for all $0 \le h \le 1$ and $\ket{G}$ is always nondegenerate. For every value of $\gamma$ the ground state $\ket{G}$ becomes fully factorized, i.e. product of single-site states, at $h_f = \sqrt{1-\gamma^2}$~\cite{GiampaoloAdessoIlluminati2008,GiampaoloAdessoIlluminati2009,
GiampaoloAdessoIlluminati2010}.
The factorizing field $h_f$ is thus always below the critical
field $h_c = 1$ and coincides with it only in the isotropic and gapless case of the $XX$ model.

\begin{figure}
\includegraphics[width=8.cm]{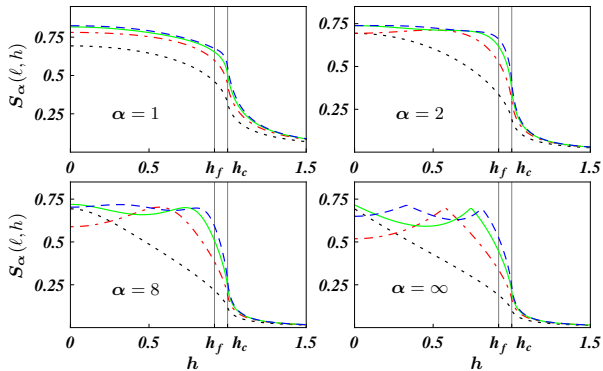}
\caption{RE $S_{\alpha}$ as functions of the transverse field $h$ for different block sizes $\ell$ in a bipartite, infinite $XY$ chain, with anisotropy $\gamma=0.4$. Dotted black line: $\ell=1$; dot-dashed red line: $\ell=2$; solid green line: $\ell=3$; dashed blue line: $\ell=4$. From the top left, clockwise, graphs of $S_1$, $S_2$, $S_8$, and $S_{\infty}$.}
\label{Figxyfunh}
\end{figure}

In Fig.~\ref{Figxyfunh} we report the RE $S_\alpha(\ell,h)$ of parity preserving ground states
as functions of $h$ for different values of $\ell$.
The von Neumann entropy $S_{1}(\ell,h)$ is monotonically decreasing both in $h$ and in $\ell$. For $1 < \alpha \leq 2$ and $h < h_f$ the RE violate monotonicity in $h$ but remain monotonically non-decreasing in $\ell$: $S_{\alpha} (\ell + 1,h) > S_{\alpha} (\ell,h)$.
Finally, for $\alpha > 2$ and $h < h_f$ all RE oscillate strongly in $h$, so that the monotonic scaling $S_{\alpha} (\ell + 1,h) > S_{\alpha} (\ell,h)$ is violated. The number and amplitude of the oscillations become more pronounced, respectively, as $\ell$ and $\alpha$ increase. The oscillations are due to a series of crossovers between the eigenvectors corresponding to the largest eigenvalue $\lambda_{max}$ of the reduced block density matrix. The number of crossovers is an increasing function of $\ell$, while the amplitude is an increasing function of $\alpha$. This phenomenon is due to the increasing weight of $\lambda_{max}$ for increasing $\alpha$. In particular, for $\ell = 2$ the maximum of $S_{\infty}$ is a parity inversion point at which the eigenvectors corresponding to $\lambda_{max}$ swap between two Bell states of opposite parity.

\begin{figure}
\includegraphics[width=8.cm]{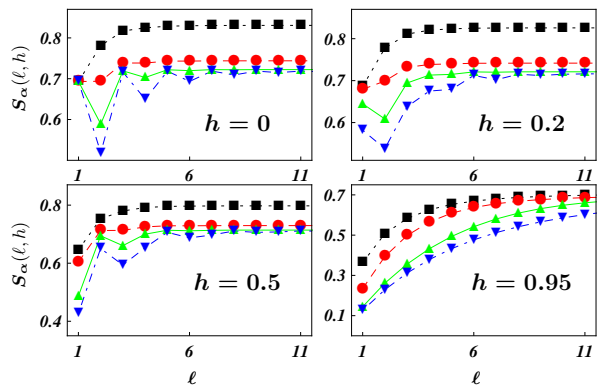}
\caption{Scaling of the RE in $\ell$ for different $\alpha$ and $h$, with $\gamma=0.4$. Squares: $\alpha = 1$; circles: $\alpha=2$; up-triangles: $\alpha=8$; down-triangles: $\alpha = \infty$. The first three panels (clockwise from top left) show the non monotonic behavior of $S^\alpha(\ell,h)$ for $\alpha > 2$ and three different values of $h$ below the factorization point $h_f \simeq 0.916$. The fourth panel illustrates the restored monotonic behavior at $h = 0.95 > h_f$, i.e. still below criticality but above factorization.}
\label{Figxyfunm}
\end{figure}

\begin{figure}
\includegraphics[width=7.cm]{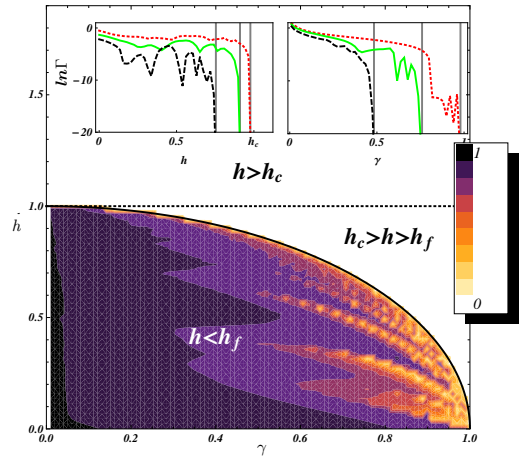}
\caption{Nonlocal order parameter $\Gamma$ as a function of $h$ and $\gamma$. Left inset: one-dimensional projection as a function of $h$ for different values of $\gamma$. Red dotted line: $\gamma=0.2$; green solid line: $\gamma = 0.4$; black dashed line: $\gamma = 0.65$. Right inset: one-dimensional projection as a a function of $\gamma$ for different values of $h$. Red dotted line: $h=0.2$; green solid line: $h = 0.65$; black dashed line: $h = 0.87$. In both insets the vertical grid lines correspond to the factorization points.}
\label{OPFig}
\end{figure}

In Fig.~\ref{Figxyfunm} we report the RE as functions of $\ell$ for different values of $\alpha$ and $h$. The violation of the area law behavior for $h < h_f$ and $\alpha > 2$ is characterized by damped oscillations in $\ell$.
The implications of the anomalous scalings reported in Fig.~\ref{Figxyfunm} can be understood by recalling that the bipartite single-copy entanglement $S_{\infty}$ is monotonic in the bipartite geometric entanglement $\mathcal{E}_G^{(2)} = 1 - \lambda_{max}$ \cite{Blasoneetal2008}. Given a system of total size $N$, $\mathcal{E}_G^{(2)}$ is the minimum distance between the ground state $\ket{G}$ and the set of pure bi-separable states $\ket{\psi_{\ell}} \; \otimes \! \ket{\phi_{N - \ell}}$, where $\ket{\psi_{\ell}}$ is a state of a block $\ell$, and $\ket{\phi_{N - \ell}}$ is a state of the remainder of the chain. From the last panel of Fig.~\ref{Figxyfunm} one sees that above factorization $\mathcal{E}_G^{(2)}$ increases monotonically in $\ell$. On the contrary, when $h < h_f$, Fig.~\ref{Figxyfunm} shows that $\mathcal{E}_G^{(2)}$ and all the RE of order $\alpha > 2$ oscillate as a function of $\ell$, with the amplitude of the damped oscillations increasing as $h$ decreases. In particular, at sufficiently small fields the true ground state $\ket{G}$ is closest to the bi-separable state $\ket{\psi_{2}} \; \otimes \! \ket{\phi_{N - 2}}$. This fact, together with the singlet-triplet inversion of the eigenvectors associated to $\lambda_{max}$, implies that below factorization $\ket{G}$ tends to order in two-spin domains. The existence of this quasi-dimerization can be explained by looking at the behavior of the RE in systems with fully dimerized ground states such as the Majumdar-Ghosh chain~\cite{Majumdar-Ghosh1969} and models with long-distance entanglement~\cite{Campos2007,Giampaolo2010}.
In these chains, if the block boundary cuts a dimer, all the resulting block RE are strongly enhanced; viceversa, if the boundary falls between two different dimers, they are strongly suppressed. By increasing the block size $\ell$ the two situations alternate themselves, giving rise to a permanent, undamped, oscillatory behavior of the block RE (including the von Neumann entropy) as functions of $\ell$.
Comparing with gapped spin models in zero field (see Fig.~\ref{Figxyfunm}, first panel) we observe an important analogy, i.e. the oscillatory behavior and the ensuing tendency to an arrangement in two-spin domains. However, the oscillatory behavior in gapped models holds only for the RE of order $\alpha > 2$ and therefore the tendency to dimerization is only partial, as it involves exclusively the eigenstates of the block reduced density matrix associated to $\lambda_{max}$. For finite $h$ (see Fig.~\ref{Figxyfunm}, second and third panels), the frequency of the oscillations decreases and, crossing factorization, the onset of a unique domain comprising the entire system restores the area-law monotonic scaling (see Fig.~\ref{Figxyfunm}, fourth panel). Exactly at factorization ($h = h_f$) the system undergoes a further phase change beyond the quantum phase transition occurring at the critical point $h_c$; in the region $h < h_f < h_c$ an entanglement-driven order is established that is superimposed on the global magnetic one. As the oscillatory behavior is maximized by the single-copy entanglement $S_{\infty}(\ell,h)$, this global order due solely to entanglement is naturally characterized by the nonlocal order parameter defined by summing $S_{\infty}(\ell,h)$ over all block sizes:


\begin{equation}
\Gamma = - \sum_{\ell=2}^{\infty}
\min [ S_{\infty}(\ell,h) - S_{\infty}(\ell - 1,h), 0 ] \; .
\label{OP}
\end{equation}
In Fig.~\ref{OPFig} we report the contour plot of $\Gamma$ for the class of $XY$ models. It vanishes identically for $h > h_f$. Approaching the Ising limit it is strongly reduced, corresponding to strongly damped oscillations. Approaching the $XX$ limit $\Gamma$ increases and diverges exactly at $\gamma = 0$, consistently with the fact that in gapless models below the critical point the oscillations persist indefinitely at any $\ell$.
Eq.~(\ref{OP}) and Fig.~\ref{OPFig} show that such nonlocal order below factorization is exclusively due to the entanglement properties of each individual ground state and therefore, unlike magnetic order, it is not a consequence of ground-state degeneracy and symmetry breaking.

We will now show that the entanglement-induced order extends to the general class of gapped $XYZ$ models:
\begin{equation}\label{Hamiltonian2}
H_{xyz} = \frac{1}{2}\sum_{i,l} J_x \sigma_{i}^x \sigma_{l}^x + J_y \sigma_{i}^y \sigma_{l}^y + J_z \sigma_{i}^z \sigma_{l}^z - h\sum_{i} \sigma_{i}^z \, .
\end{equation}
Here $J_\mu$ are the spin-spin couplings along the $\mu=x,y,z$ directions and, without loss of generality, we set $J_x=1\ge |J_y|,|J_z|$. These systems undergo a quantum phase transition at $h = h_c$. The relevant difference between the general $XYZ$ models and the $XY$ chain lies in the fact that while the ground state of the latter is always subject to a factorization at $h = h_f$, the models described by Eq.~(\ref{Hamiltonian2}) admit a factorized ground state at a finite value $h = h_f =\sqrt{(J_x+J_z)(J_y+J_z)}$ if and only if $J_z \ge -J_y$ ~\cite{GiampaoloAdessoIlluminati2008,GiampaoloAdessoIlluminati2009}.
The models of the class Eq.~(\ref{Hamiltonian2}) are not exactly solvable. In order to determine the block reduced density matrices we diagonalized the system by means of the Density Matrix Renormalization Group (DMRG)~\cite{revscholl,JCP} applied to open chains of up to $128$ spins.
For models admitting a factorization point, the results of the $XY$ case carry over essentially unmodified to the general $XYZ$ instance.
\begin{figure}
\includegraphics[width=7.cm]{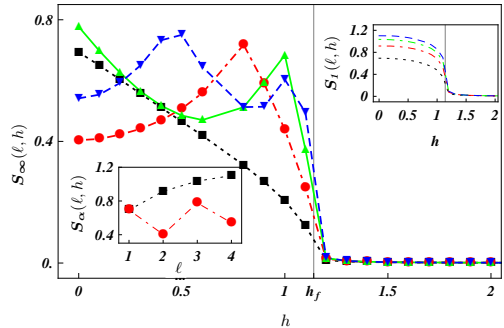}
\caption{Single-copy entanglement $S_{\infty}$ as a function of $h$ for different values of $\ell$ in a $XYZ$ model with $J_x=1$, $J_y=0.7$, $J_z=0.3$, and $h_f \simeq 1.14$. Squares: $\ell = 1$; circles: $\ell = 2$; up-triangles: $\ell = 3$; down-triangles: $\ell = 4$. Upper right inset: von Neumann entropy $S_1$. Lower left inset: $S_{\infty}$ (circles) and $S_1$ (squares) as functions of $\ell$ at $h=0$.}
\label{Figwith}
\end{figure}
In Fig.~\ref{Figwith} we report the scaling of the RE in $XYZ$ models that possess a factorizing field $h_f$ ($J_z \ge -J_y$). The qualitative behavior is analogous to that of the $XY$ models and confirms the existence of an entanglement-induced order of two-spin domains for fields $h < h_f$ also in the general case.
\begin{figure}
\includegraphics[width=7.cm]{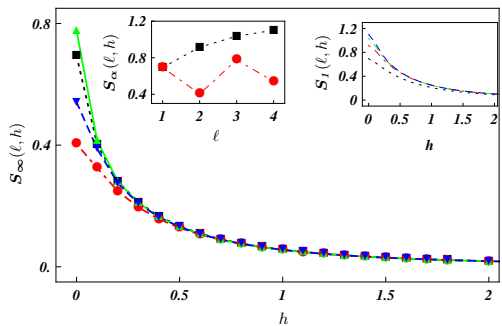}
\caption{Single-copy entanglement $S_{\infty}$ as a function of $h$ for different values of $\ell$ in a $XYZ$ model with $J_x=1$, $J_y=-0.7$, $J_z=-0.3$, and no factorization point. Squares: $\ell = 1$; circles: $\ell = 2$; up-triangles: $\ell = 3$; down-triangles: $\ell = 4$. Upper right inset: von Neumann entropy $S_1$. Upper left inset: $S_{\infty}$ (circles) and $S_1$ (squares) as functions of $\ell$ at $h=0$.}
\label{Figwithout}
\end{figure}
The investigation of models that do not admit a factorized ground state at any finite value of the magnetic field is summarized in Fig.~\ref{Figwithout}.
At variance with the case admitting ground state factorization, the area-law order between blocks of different size $S_{\alpha} (\ell + 1,h) > S_{\alpha} (\ell,h)$ appears to be \emph{always} violated for \emph{all} values of $h$, as one can see comparing the main plots of Figs.~\ref{Figwith} and ~\ref{Figwithout}. On the other hand, a rigorous confirmation of this behavior for all values of $h$ is beyond the current numerical possibilities, since the differences between RE of the same order $\alpha$ and different block size $\ell$ fall off extremely rapidly with $h$. In the absence of factorization all the RE behave smoothly and do not acquire local maxima. Moreover, considered as functions of the block size $\ell$ (at fixed external field $h$) they exhibit the same oscillating behavior as in models with factorized ground states (See the left inset of Fig.~\ref{Figwithout}). These features can be intuitively understood by considering that in these models factorization, so to speak, is moved towards infinitely large values of $h$, corresponding to a classical saturation. This behavior confirms that the entanglement-induced order can exist also in a symmetry unbroken phase with vanishing local order parameters.

In conclusion, we have investigated the R\'enyi entanglement entropies in one-dimensional gapped quantum spin models. We have showed that a violation of the area law scaling behavior, analogous to that occurring in the gapless cases, holds for all R\'enyi entropies of order $\alpha > 2$ and for external fields $h < h_f < h_c$ for models admitting ground-state factorization. For models that do not admit factorization, it appears to extend to all values of $h$. This anomalous scaling characterizes the existence of a novel type of entanglement-driven order. Unlike the magnetic order, this ordering does not rely on the onset of ground-state degeneracy, but is rather due to the entanglement properties of each parity preserving ground state. The factorizing field plays a key role in the understanding of the anomalous scaling of the RE, by clarifying the origin of this phenomenon even in the gapless case in which the factorization and critical points coincide. It also explains why the anomalous scaling is not observed in systems, like the Ising model, for which $h_f = 0$. The maximal violation of the area law occurs in the case of the RE of order $\alpha = \infty$, the single-copy entanglement. This fact allows to introduce a nonlocal order parameter, defined as the sum of the single-copy entanglement over all block sizes, that is nonvanishing for all fields $h < h_f$ and vanishes identically for $h > h_f$. On the other hand, the single-copy entanglement is monotonic in the bipartite geometric entanglement $\mathcal{E}_G^{(2)}$, which has been shown to be a universal lower bound to ground-state frustration \cite{Giampaolo2011}. This correspondence suggests the existence of an intimate relation between entanglement scaling and frustration of purely quantum origin that will be the subject of further analysis. In such a perspective, the present investigation might be fruitfully extended to geometrically frustrated models and systems with topological order and fractional excitations.

{\em Acknowledgements:} One of us (SM) thanks M.~Rizzi for discussions. SMG and FI acknowledge support from the EU STREP Projects HIP, Grant Agreement No. 221889, and iQIT, Grant Agreement No. 270843. SM  acknowledges the PwP project (www.dmrg.it) for the DMRG code, the DFG (SFB/TRR~21), the bwGRiD, the EU IP Project AQUTE, Grant Agreement No. 247687, and the EU STREP Project PICC, Grant Agreement No. 249958, for support.

\end{document}